\documentclass[12pt]{article}
\usepackage{epsfig}
\usepackage{amsfonts}
\usepackage{amssymb}
\usepackage{amsmath}
\usepackage{amsthm}
\usepackage{latexsym}
\usepackage{graphicx}
\usepackage{bm}
\usepackage{tabularx}
\usepackage{tikz}
\usepackage{booktabs}
\usepackage{enumerate}
\usepackage{caption}
\usepackage[titletoc]{appendix}
\usepackage[numbers]{natbib}
\usepackage{float}
\usepackage{comment}
\usepackage{tabularx}
\usepackage{fancyhdr}
\usepackage{epstopdf}
\usepackage{subcaption}
\usepackage{dsfont}

\catcode`\@=11

\newcommand{\PP}[1]{\mathrm{P}\left( #1 \right)}

\usepackage[pdftitle={The Spotify Top 200},
  pdfauthor={CORAL},
  pdffitwindow=true,
  breaklinks=true,
  colorlinks=true,
  urlcolor=blue,
  linkcolor=red,
  citecolor=red,
  anchorcolor=red]{hyperref}

\numberwithin{equation}{section}
\numberwithin{table}{section}
\numberwithin{figure}{section}

\theoremstyle{plain}

\theoremstyle{definition}

\theoremstyle{remark}

\begin{document}

\title{Analyzing the Spotify Top 200 Through a Point Process Lens}

\author{Michelangelo Harris, Brian Liu,  Cean Park, Ravi Ramireddy,\\ Gloria Ren, Max Ren, Shangdi Yu, Andrew Daw, and Jamol Pender\footnote{Correspondence: \texttt{jjp274@cornell.edu}; 228 Rhodes Hall, Cornell University, Ithaca, NY, 14853.}
\\
\\
School of Operations Research \& Information Engineering
\\
Cornell University, Ithaca, NY}

\maketitle

%\tableofcontents
\abstract{
\textit{Every generation throws a hero up the pop charts.} For the current generation, one of the most relevant pop charts is the Spotify Top 200. Spotify is the world's largest music streaming service and the Top 200 is a daily list of the platform's 200 most streamed songs.  In this paper, we analyze a data set collected from over 20 months of these rankings. Via exploratory data analysis, we investigate the popularity, rarity, and longevity of songs on the Top 200 and we construct a stochastic process model for the daily streaming counts that draws upon ideas from stochastic intensity point processes and marked point processes. Using the parameters of this model as estimated from the Top 200 data, we apply a clustering algorithm to identify songs with similar features and performance.
}\\


\section{Introduction}

Streaming music, or more accurately streaming audio, is a way of delivering sound without requiring you to download files from the internet. Spotify is the world's largest music streaming service, currently available as both desktop and mobile applications. Spotify provides free streaming of music to its users, with an option for a paid premium membership with added benefits such as no advertisements and the ability to listen to music offline. As of April 2019, Spotify had 217 million monthly active users worldwide, 100 million of which are paid subscribers \cite{spotify2018earnings}. On top of Spotify itself growing, streaming has become increasingly central to the music industry. In 2018, streaming accounted for 77\% of all music consumption relative to 23\% combined for physical and digital sales, compared to 66.1\% and 33.9\%, respectively, in 2017 per the industry report provided by \citet{buzzangle2018year}. Streaming's music industry dominance is already influencing the structure of new releases, such as in the influx of very short (e.g. $\leq$ 2.5 min.) singles \cite{pearce2018considering,quartzy2018reason} or in the recent popularity of \textit{extra} long playing (e.g. $\geq$ 80 min.) albums \cite{mapes2016against}. As these trends continue to manifest themselves throughout popular music, it is both increasingly relevant and increasingly valuable to understand how music evolves through the streaming ecosystem.

This paper aims to begin that expedition through an analysis of the Spotify Top 200, a chart of the 200 most frequently streamed songs each day on Spotify. This investigation is built on a data set scraped from over 20 months of the United States based edition of these rankings, spanning from 2017 to 2018. To put this time capsule in a pop music context, this is a particular period of dominance for Drake that may eventually be viewed as his peak, at least commercially. He has claim to 6 of the 32 number one songs in this data set and this includes the three highest recorded daily play counts, each of which is over 4.5 million streams. Moreover, on the day his 2018 album \textit{Scorpion} was released, all 25 tracks were in the top 27 most streamed songs, including places 1 through 23. Hip hop is known to be the most popular genre currently on streaming services in the U.S. \cite{buzzangle2018year}, and that fact is reflected throughout this data and the songs discussed herein.

One key insight we make from this data is that a song's daily stream count can be modeled as discrete observations of a non-stationary point process, specifically a marked or stochastic intensity point process \cite{sigman2019marked}. While this construction is entirely motivated from data, we can note that there are conceptual similarities between music streaming services and the theoretical details of several well-known point processes of this type.  For example, an oft-recurring theme of this data analysis is the influence that an external happenings can have on the streams a particular song receives. As we will demonstrate, events like album releases, music video release, and public performances can lead to significant surges in the song's stream count. This type of exogenous excitement closely resembles the type of jumps in the stochastic intensities of Cox processes \cite{daley2003introduction}. Additionally, the fact that the Spotify Top 200 chart is also a playlist available to all users on the service means that as a song climbs higher and higher in rank it becomes more and more visible. Hence a song that is heavily streamed one day may be more likely to again be heavily streamed again the next day, yielding a discrete time form of \textit{self-excitement}, see \cite{hawkes1971spectra,seol2015limit}.  This also captures the social contagion of a song's popularity, and this epidemic-like information sharing has recently been formally connected to self-exciting processes as well \cite{rizoiu2018sir,daw2019queue}. Thus we can see that streaming services should experience a combination of both externally driven and self-excited behavior. Processes of this sort have been studied in the literature as dynamic contagion processes \cite{dassios2011dynamic}. Using the point process model for this streaming system, we will uncover connections between songs through a clustering analysis, thus generating further insight into the Top 200 data.
%\textcolor{blue}{Maybe we also cite the new paper on Matryoshkan since it actually would provide the moments a process like this.}

A quick look at the current pop charts shows that these dynamics are not unique to the time of this data set. Rather, recent pop music storylines seem to indicate that public events are becoming even more important for commercial success. Perhaps this is most obviously championed by Lil Nas X and his masterful manipulation of publicity in the record-breaking 19 weeks ``Old Town Road'' spent atop the Billboard Hot 100 chart, all of which occurred after this data set concludes. Beyond just catching lightning in a bottle, music critics have noted that sustained success of ``Old Town Road'' can be at least partially traced to the expertly timed sequence of remixes, videos, and even internet memes that maintained public interest \cite{breihan2019old}. And while ``Old Town Road'' is of course the most dominant example of these strategies so far, it is not the only one. For example, Billie Eilish buoyed her eventual number one hit ``bad guy'' by releasing a remix featuring Justin Bieber. As an example of the effect of an external event not directly controlled by the artist, consider Lizzo's Billboard topping song ``Truth Hurts.'' Originally released in 2017, the Minnesotan singer/rapper/flutist's sleeper hit surged in popularity after being featured in the 2019 Netflix film \textit{Someone Great}. It is also worth noting that both ``Truth Hurts'' and ``Old Town Road'' experienced an initial viral growth through use in videos shared on the app TikTok, which then led to success on music streaming platforms like Spotify. Based on observations like these, the goal of this paper is to explore, model, and understand the dynamics of a pop song on streaming platforms, and this investigation is centered around the Spotify Top 200 data set.

%should cite dynamic contagion - self-excitement and external excitement are both present)

\subsection{Description of Data}

The data used in this paper is a daily record of the Top 200 playlist on Spotify, which contains the top 200 most streamed songs on that day. As defined by Spotify, one day spans from 3:00 PM UTC through 2:59 PM UTC on the next. We have scraped this data set from the publicly available Spotify Top 200 charts, specifically from the U.S. based rankings. The code to do so is derived from open source code used to form a Kaggle data set containing the worldwide Top 200 rankings for all of 2017 \cite{kaggle2017}. We have collected the streaming data from this source over the date range January 1, 2017 to September 12, 2018, but the code used to scrape the data can easily by recompiled to acquire up to current records. The fields contained in this data are date, position (rank), song name, artist, and the number of streams of that song on that date. Spotify counts a ``stream'' for a song after a user has listened for 30 seconds \cite{Spotify-qa}. The daily stream count is then the total count of these events across the time period.

\subsection{Contributions \& Organization}
The questions we have answered about this data set are focused around how stream counts vary by position in the Top 200 and over time for particular songs. This data set motivates a point process model of a song's jump and decay in streams, which in turn enables further model-based analysis. This investigation is organized throughout the remainder of this paper as follows:
\begin{itemize}
\item In Section~\ref{secExplore}, we perform an exploratory data analysis for the Spotify Top 200. Our analysis include the exclusivity of each ranking position, the duration of a song's time on the charts, and close inspection of the behavior of songs at the highest and lowest positions of the rankings.
\item In Section~\ref{secModel}, we take these findings as motivations to construct a stochastic model for a song's daily streaming totals. This stochastic process model is a generalized Poisson process, and we describe how to obtain its parameters from data via maximum likelihood estimation. We also explain how to use linear regression to estimate the single parameter in a special, parsimonious sub-case of the model.
\item In Section~\ref{secCluster}, we use the model from Section~\ref{secModel} to obtain additional insights into the Top 200 data set. We inspect both the strengths and weaknesses of this model in representing this data set. Based on these observations, we perform a clustering analysis on the songs contained in this data set using only the estimated model parameters and an empirical measure of their fit. This leads us to insights that appear to transcend both the model parameters and the data itself, recognizing connections that were not obvious until after the data's time range.
\end{itemize}

%Questions raised by our EDA include: Can a simple (parsimonious) point process model usefully capture the complex operational reality of the dynamics of the Spotify Top 200?  How do songs in the Top 200 decay over time? Do the decay rates depend on the genre or the rank of the song?  How much money does an artist make conditional on rank? The significance of such questions and our related findings raise the need for novel point process models and theory, which we present here as opportunities for future research.
%Other import questions include:
%How do songs flow out of the top 200?
%How many streams per user for the Top song?
%What is the distribution of streams for the Top song?
%Our goal is to explore the Spotify data through a point-process lens.  Through exploratory data analysis (EDA), we find that we can explain the dynamics of the streaming process of each rank by a non-stationary point process with an exponentially decaying kernel.  Moreover, the average number of streams by rank follows a power law distribution.

\section{Exploratory Analysis of the Spotify Top 200}\label{secExplore}

We begin our analysis by exploring this Spotify data set and examining the relationships between streams, songs, and positions on the Top 200 chart. In Subsection~\ref{subsecTop200Plots} we take a position-based perspective and investigate the popularity and the rarity associated with each spot on the Top 200 chart. These leads us to consider the lifespan of each song, and in Subsection~\ref{subsecDuration} we measure the frequency and duration of the songs that made it on this list. Then in the final two subsections we contrast the biggest hits with those that barely made the cut, with Subsection~\ref{subsecNumberOnes} devoted to the number ones and Subsection~\ref{subsecIrrelevant} focused on the last place of the Top 200.

\subsection{The Prestige and Intrigue of the Top 200}\label{subsecTop200Plots}

\begin{figure}[t]
            \centering
             \includegraphics[width = .9\textwidth]{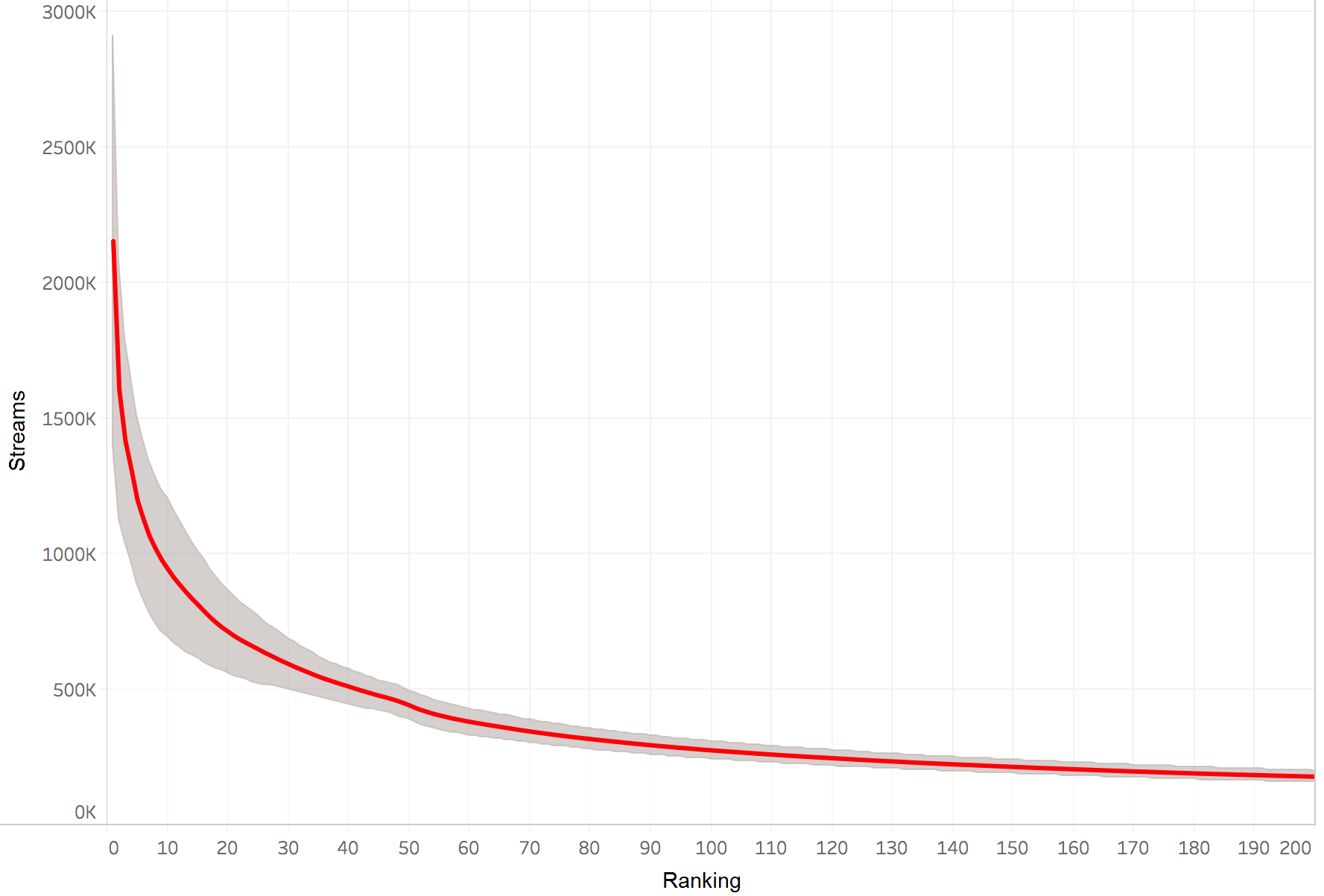}
             \caption{The average number of daily streams received at each position in the Spotify Top 200, plus and minus one standard deviation.}
            \label{fig:avg_stream_pos}
%            \includegraphics[width = 0.9\textwidth]{stdev_stream_pos.png}
%            \label{fig:stdev_stream_pos}
        \end{figure}

In viewing the Spotify Top 200 data set as a daily recording of a list of 200 tracks and their associated play counts, it is natural to wonder how the places in these rankings compare to one another as measured in terms of these songs and streams. For example, by definition the first-placed song in the Top 200 is played more often than the second-placed song (or any other), but how much more? In Figure \ref{fig:avg_stream_pos}, we plot the average across time of the number of daily streams received in each of the 200 ranks. The shaded gray area represents a standard deviation above and below the average (red) line.

       One can observe that the decrease in the number of streams is most extreme for the top positions, meaning that the difference in streams between the top song and the runner-up is much larger than the difference between streams in the $199^\text{th}$ and $200^\text{th}$ positions. This effect can perhaps be even more easily observed when comparing the difference between ranks 1 and 10 with the difference ranks 191 and 200. One can see that the top spot averages over 2 million streams each day whereas the tenth spot receives approximately 900K listens, yet the change from 191 to 200 is hardly noticeable. This observation can be formalized, as fitting a \textit{power law} curve of the form $f(x) = a x^{-b}$ to the data yields coefficients $a = 2.3689 \times 10^6$ and $b = 0.5426$ with an $R^2$ value of 0.9805. The power law tail of the average streams chart may point to a connection between Spotify and other social platforms, since the power law distribution is closely associated with connectivity in human social networks, see e.g. Chapter 18 of \citet{easley2010networks} for an overview. Furthermore, this power law structure confirms the observed ``diminishing returns'' effect for song rank, since this function dictates that a song would need exponentially more streams to move from rank 2 to 1 than from 200 to 199.

       Reasoning about this chart leads us to two intuitive explanations for this behavior. First, the prestige associated with the highest rankings  naturally suggests that only the biggest hits can achieve the massive number of daily streams needed to make it to the top of the list, as it is certainly more impressive to be the \#1 song than it is to be \#10 song. By comparison, the $191^\text{st}$ ranking does not feel particularly more commendable than the $200^\text{th}$. Second, one can also observe a natural intrigue associated with the top of the list: listeners looking for a new hit song may be much more likely to look at the top of the list than at the bottom. This yields a {self-excitement} effect, as the Top 200 is both a reflection of popularity and a conductor of it. That is, a song receiving a large number of streams suggests that it is likely to receive even more as a consequence of its increased exposure on the Top 200 playlist. This power-law structure also leads us to interesting observations on the behavior of the bottom end of the rankings. The slow decay of streams between the lower ranked songs hints at the idea of the \textit{long tail} first written about in \citet{anderson2004long}. As songs decrease in the rank of their popularity their actual play-count does not as noticeably taper offer. \citet{anderson2004long} predicted that this would be a hallmark of the more niche-based entertainment offerings the internet could provide. Thus, as a potential direction of future study given additional access to data, it would be interesting to see how this extends beyond the inherent popular Top 200 into the worlds of independent and experimental music.

        %Interestingly enough, this distribution stream standard deviation suggests that the number of streams alone does not account for a song's position. High volatility in song positions 1 through 30 may mean that there exists a type of seasonality, or third variable that allows songs to reach the top of the charts.

\begin{figure}[t]
    \centering
    \includegraphics[width = \textwidth]{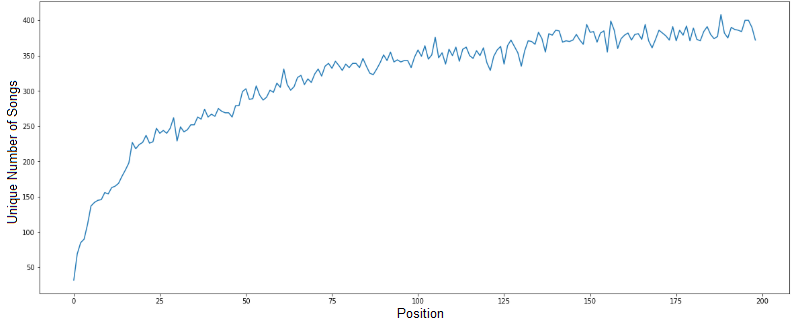}
    \caption{The number of unique songs observed in the data at each position in the Top 200.}
    \label{fig:unique_pos}
\end{figure}

        In addition to considering the number of streams associated with each position in the Spotify Top 200, we are also interested in the number of songs connected to each of these ranks. Hence, in Figure \ref{fig:unique_pos} we plot the number of distinct songs observed at each rank in this chart over the course of this data set, which spans 620 days from January 1, 2017 to September 12, 2018.  As one might expect based on the dramatic decrease in popularity at each position in Figure~\ref{fig:avg_stream_pos}, we can observe in Figure~\ref{fig:unique_pos} that it is also increasingly rare to see a song achieve a higher and higher rank. For example, less than 50 songs ever reached the top, whereas there were approximately 400 unique songs observed at each of the bottom 50 positions.

    The increased rarity of the top of these charts naturally leads one to wonder about the life span of songs in this data set. In the next subsection, we will investigate both how long a song stays on the Top 200 after it first appears and how this duration changes based on the position the songs achieves.

% We see a dramatic increase from position 1 - 25, and then a gradual leveling off. This observation supports the hypothesis that songs that achieve a higher rank will stay for longer periods of time than those that have a ranking closer to to the bottom of the list. The average number of unique songs that are recorded at any given rank from Janurary 1 2017 to September 12 2018 is 319.5.

% Additional questions we wish to explore include: when do rank 1 songs have the highest number of streams, and why? What additional factors or third variables may account for the large deviation in average streams for the highest ranks?

\subsection{15 Seconds of Fame? Quantifying the Duration of a Pop Hit}\label{subsecDuration}

\begin{figure}[t]
    \centering
    \includegraphics[width = .9\textwidth]{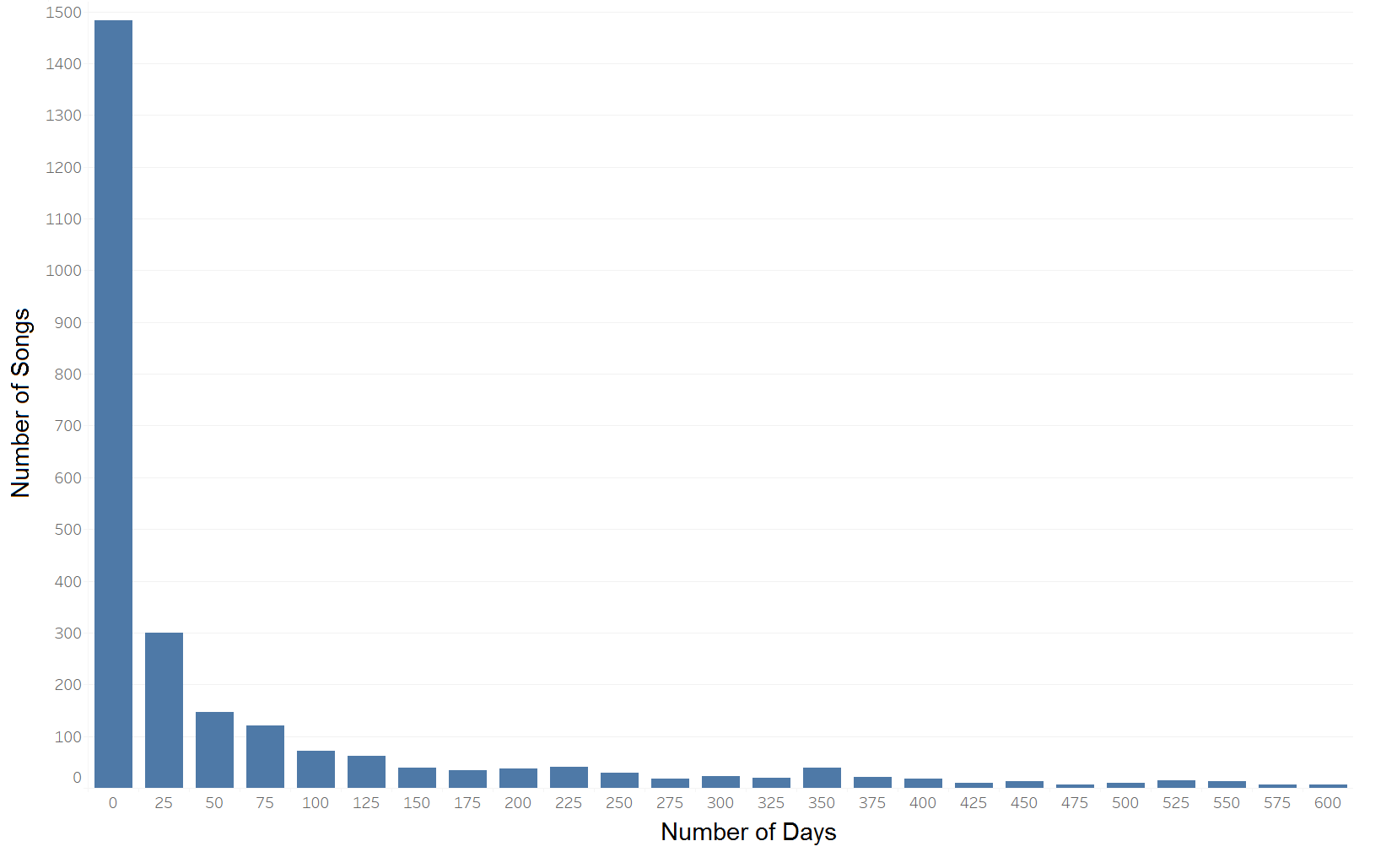}
    \caption{Histogram of the first duration of each song appeared in the data set.}
    \label{fig:firstlife25}
\end{figure}

In order to obtain a better understanding of the life cycle of a song within the Spotify Top 200, let us define the first life of a given song as the number of consecutive days from when a song first entered the Top 200 until it no long appears on the charts. Before proceeding with the analysis, it is important to note that by the nature of this data set some songs have longer first lives than what we can calculate because they may have initially appeared in the charts before January 1, 2017. Nevertheless, this investigation still gives us an idea of the duration of a song on the Top 200. In Figure~\ref{fig:firstlife25} we first plot the frequency histogram of the length of the first lives of all songs in the data. That is, we define the first life of a given song as the number of consecutive days from when a song first entered the Top 200 until it no long appears on the charts.

As can be observed, the largest collection is for the shortest duration: nearly 1,500 songs have a duration of less than 25 days. Still, there are 9 songs that have durations for over 500 days in this 620 day data set. For reference, the titles of these songs and their recorded durations are given below in Table~\ref{table9Longest}. To further specify the observed duration of songs, we also provide the percentage of songs in the data that lasted no more than a day, week, month, and year in Table~\ref{tableDurBound}. As one can see, just less than a third of the songs that appeared on the Top 200 chart only lasted a day and moreover more than two thirds of the songs did not stay on the chart for more than a week after their first appearance. The existence and rarity of the tail of the distribution in Figure~\ref{fig:firstlife25} is also demonstrated in Table~\ref{tableDurBound}, as less than 16\% of songs remained in the Top 200 for over a month, and less than 0.8\% lasted for over a year, yet from Table~\ref{table9Longest} we know that 5 songs appeared in well over 18 months of this data set. By comparison, the mean first life duration is approximately three weeks. This shows a heavy-tailed phenomenon appearing in these durations, as extreme outliers are known to occur.

\begin{table}[h]
    \centering
    \begin{tabular}{c c}
    \textbf{Name of Song}  & \textbf{First Life} \\
    \hline
    Goosebumps                                                                            & 617                             \\
    Congratulations                                                                       & 617                             \\
    Location                                                                              & 616                             \\
    Shape of You                                                                          & 603                             \\
    Believer                                                                              & 585                             \\
    XO TOUR Lif3                                                                          & 535                             \\
    HUMBLE.                                                                               & 528                             \\
    Young Dumb \& Broke                                                                   & 519                             \\
    LOVE. FEAT. ZACARI                                                                    & 512                             \\
   % Thunder                                                                               & 500
    \end{tabular}
    \caption{The 9 songs with first life duration longer than 500 days in this data set. }\label{table9Longest}
\end{table}

\begin{table}[h]
    \centering
    \begin{tabular}{c c}
    \textbf{Percentage of Songs} & \textbf{First Life Duration} \\
    \hline
    31.80\%                                                                          & $\leq$ 1 day                                              \\
    68.67\%                                                                          & $\leq$ 1 week                                             \\
    84.04\%                                                                          & $\leq$ 1 month                                           \\
    99.23\%                                                                        & $\leq$ 1 year
    \end{tabular}
    \caption{Percentage of songs that have durations bounded by a day, week, month, and year. }\label{tableDurBound}
\end{table}

\begin{figure}[t]
    \centering
    \includegraphics[width = \textwidth]{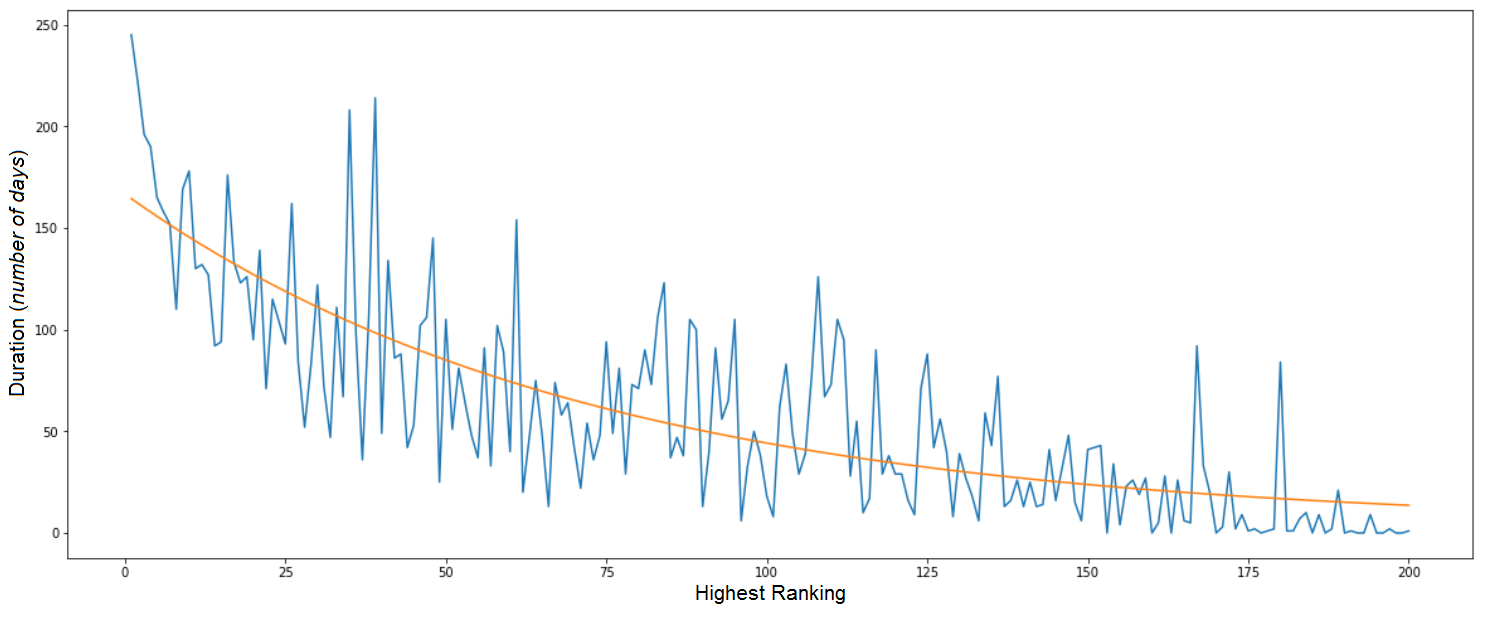}
    \caption{Average first life durations given the highest position achieved in the Top 200.}
    \label{fig:staying_time_decay}
\end{figure}

When considering the first life duration of a given song, it seems natural to wonder how this is effected by the rank of the song. In Figure~\ref{fig:staying_time_decay} we plot the empirical conditional expectation of the duration of the song given the highest ranking it achieved in the chart. Additionally, this plot contains an exponential decay curve fitted to the data. This figure shows what one might reasonably expect: the more popular songs also last longer on the charts. In particular, we can see that songs that achieved the top rank averaged a duration of nearly 250 days, whereas the songs that never cracked 190 stay for a very brief time. In the following two subsections we further contrast the top and bottom ranks of the Top 200, beginning now with a closer look at the biggest hits of all in Subsection~\ref{subsecNumberOnes}.

\subsection{Far From the Shallow Now: Examining \#1 Ranked Songs}\label{subsecNumberOnes}

\begin{figure}[t]
    \centering
    \includegraphics[width = \textwidth]{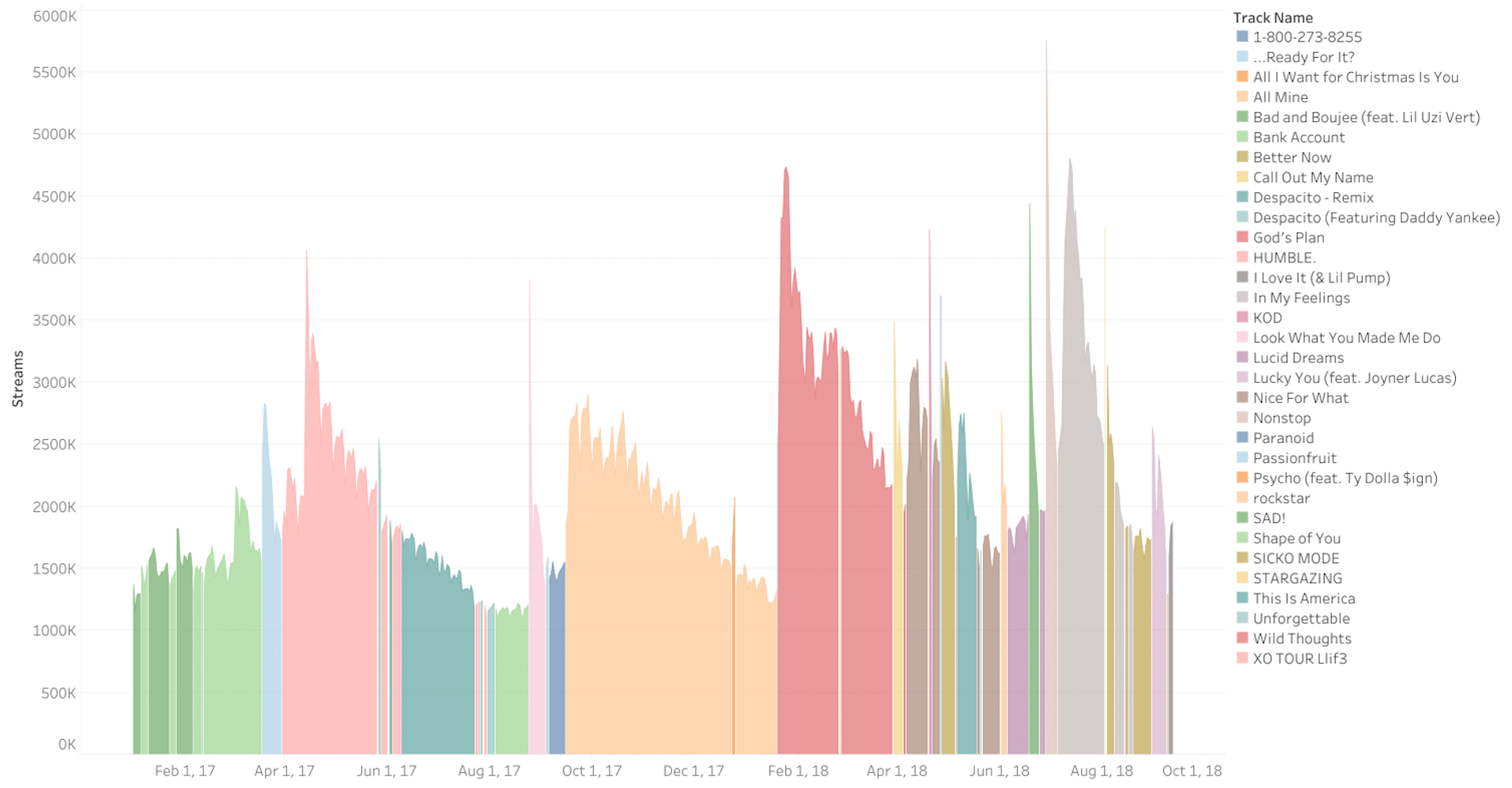}
    \caption{Top Spotify Songs from January 1 2017 to September 12 2018.}
    \label{fig:top_songs}
\end{figure}

From Subsection~\ref{subsecTop200Plots}, we know that it is very rare for a song to become the number one track on the Spotify Top 200. Figure~\ref{fig:unique_pos} showed us that only 32 songs ever reach the ultimate rank in the 620 days recorded in this data set. This matches with what he have seen in Subsection~\ref{subsecDuration}, as Figure~\ref{fig:staying_time_decay} revealed that top ranked songs have a significantly longer life span than the average song does. In this subsection we will explore these 32 hits in more detail through Figure~\ref{fig:top_songs}. This plot shows the number of streams received by the number on song throughout the full time range of the data. As the number one song changes, so does the color of the shaded curve.

We can make several interesting observations from Figure~\ref{fig:top_songs}. First, we can see that in most cases a change of the guard at the top of the Top 200 occurs when a new song jumps up to a higher number of streams. This isn't exclusive, such as the more passive takeovers seen in August 2017, but it does seem to be the primary type of transition. We can also see that the number one song's daily stream count primarily just decays after its debut at the top spot. A notable pattern of exceptions to this are subsequent jumps in streams, such as what is seen in late April 2017. Looking closer at this specific example can give us insight into the source of these dynamics. The song occupying the number one spot for April and May 2017 is ``HUMBLE.'' by Kendrick Lamar. This hit was the advance single for Lamar's 2017 album \textit{DAMN.} and was released to much fanfare and attention along with its accompanying music video, which was released simultaneously on March 30, 2017. As a single, ``HUMBLE.'' catalyzed the feverish anticipation for Lamar's third official studio recording. When the full album was then released two weeks later on April 14 the song received even more listens, which is reflected in the approximately 2 million streams added to its play count on this date.

A similar pattern can be seen in January and  February 2018 for Drake's hit single ``God's Plan,'' which eventually appeared on his June 2019 album \textit{Scorpion}.  This song was first made available to stream on January 19, 2018, at which point we can see that it jumps up to an unprecedented height in the streaming data. There is then a second, smaller jump in streams around a month later. Rather than being the eventual June album-release jump, this increase in streams coincides with the release of the music video for ``God's Plan,'' a viral visual in which Drake gave away large amounts of cash to strangers in Miami, FL. These examples give us intuition for the jump-and-decay dynamics seen in Figure~\ref{fig:top_songs}. By the nature of streaming's instant access, new songs or songs with new events can draw a rush of new listens, but this attention can fade over time, possibly in favor of newer options.

\subsection{Barely Famous: The 200$^\text{th}$ Song on the Top 200}\label{subsecIrrelevant}

By comparison to the number one song, we have seen that the 200$^\text{th}$ ranked track receives far fewer streams (see Figure~\ref{fig:avg_stream_pos}) and remains on the chart for much less time (see Figure~\ref{fig:staying_time_decay}). However from Figure~\ref{fig:avg_stream_pos}, we know that there isn't such a substantial difference in streams between the 200$^\text{th}$ song and the other tracks on the tail of the Top 200. Thus, by inspecting the behavior of the songs in the last ranked spot we can get an impression of songs that achieve modest popularity on Spotify. From Figure~\ref{fig:unique_pos}, we also know that this represents a much larger population of music than the top hits.

\begin{figure}[t]
    \centering
    \includegraphics[width = .95\textwidth]{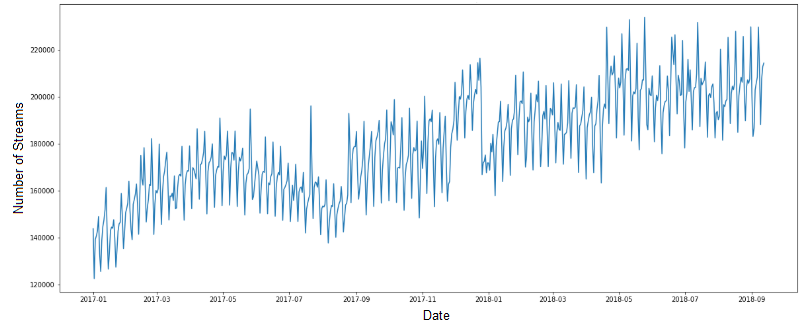}
    \caption{Number of daily streams received by the 200$^\text{th}$ ranked song across the full data set.}
    \label{fig:min_enter_pos}
\end{figure}

\begin{figure}[t]
    \centering
    \includegraphics[width = .95\textwidth]{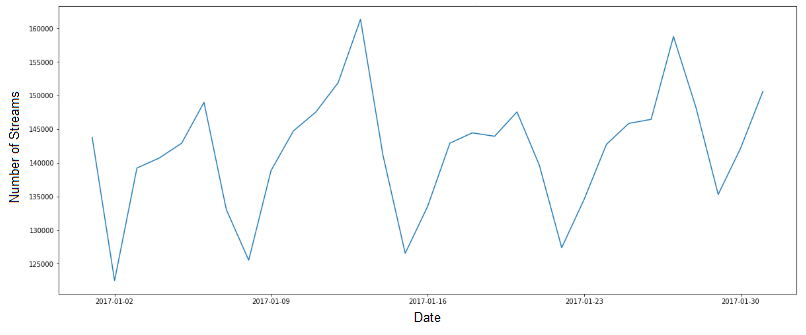}
    \caption{Number of daily streams received by the 200$^\text{th}$ ranked song in January 2017.}
    \label{fig:min_enter_pos_month}
\end{figure}

First, in Figure~\ref{fig:min_enter_pos} we plot the number of streams received by the rank 200 song across each of the 620 days in the data set. While there may be some mild long-term seasonality, the most striking features are the short-term periodicity and the overall upward trend in streams. Let us first discuss the latter. One can observe that at the beginning of 2017 the daily stream count is oscillating around 140,000 plays while at the start of September 2018 the center is closer to 200,000 streams. This appears to be a quite accurate reflection of the growth of the music streaming in the U.S. as a whole, which is reported to have increased 42\% from 2017 to 2018, per the annual industry report from \citet{buzzangle2018year}. Returning to short-term periodicity, let us now look closer at the first month of daily stream data in Figure~\ref{fig:min_enter_pos_month}.

On this time scale, the pattern becomes evident: there is clearly a weekly structure to the number of streams. Recalling that New Years Day fell on a Sunday in 2017, we can see that the number of streams peaks each week on Saturday and mostly increases throughout the week from Sunday onward. We can note multiple explanations for this. As a simple first observation, it seems natural for the weekend to be a popular time to listen to music as a form of relaxation outside of the work week. Furthermore, in a reasoning fundamentally based on the music industry, since 2015 there has been a globally recognized standard of releasing new music on Fridays. As noted in \citet{flanagan2015industry}, this was an eventual consequence of Beyonc\'e's surprise release self-titled album in 2013. \textit{Beyonc\'e} debuted online on a Friday without any advance notice, a then-unprecedented concept for a major pop release, and was massively successful. In her own words, Beyonc\'e ``changed the game with that digital drop.''\footnote{From ``Feeling Myself'' by Nicki Minaj feat. Beyonc\'e, see \texttt{https://genius.com/4532100}.} Previously, the weekly release date varied by country and shaped each nation's weekly sales and streaming measurements accordingly. The new global standard of a Friday release date is thought to be better catered to the culture of instant music access foreshadowed \textit{Beyonc\'e} and now associated with online streaming platforms, and Figure~\ref{fig:min_enter_pos_month} shows how this has been manifested on Spotify. It is worth noting that despite this influence her music has had, Beyonc\'e the artist is almost entirely absent from our data exploration. This is most likely due to the fact that at the time her most recently released album \textit{Lemonade} was only available on the competing platform TIDAL, of which she is a partial owner.

\section{Modeling Streaming through Stochastic Processes}\label{secModel}

\begin{figure}[t]
\centering
\includegraphics[width=\textwidth]{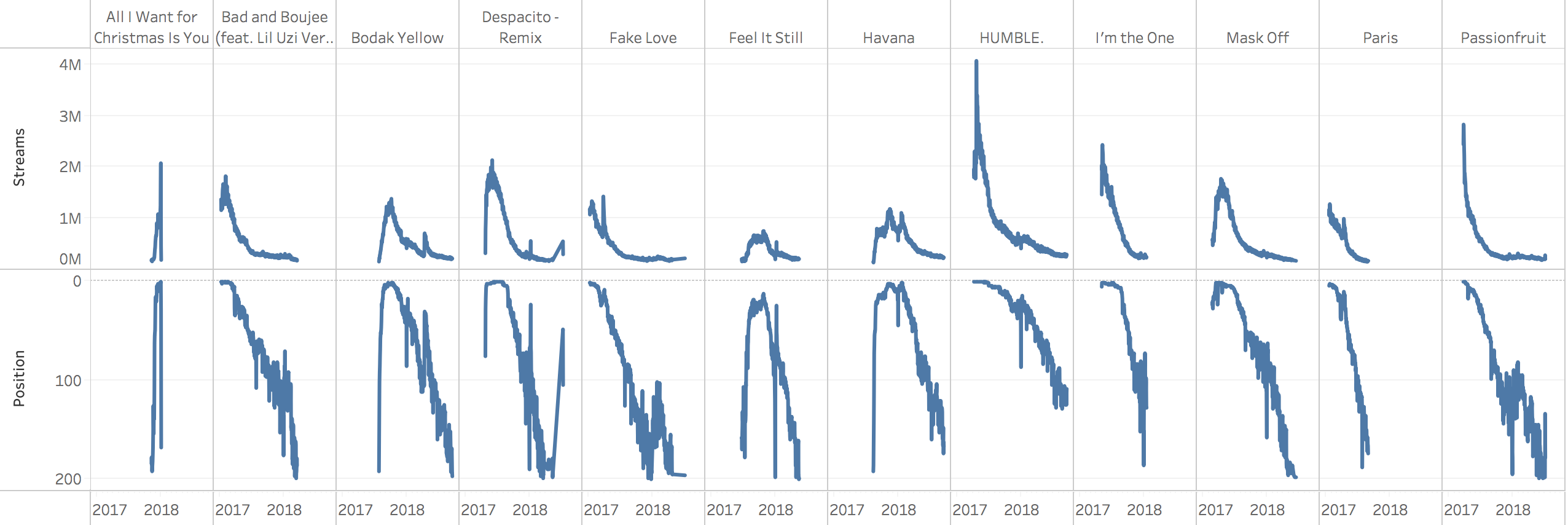}
\caption{Sample paths for 12 example songs in the Spotify Top 200 data set.}\label{fig:sample_paths}
\end{figure}

Following the observations we have made throughout our exploratory data analysis in Section~\ref{secExplore}, we now propose a stochastic process model for the times that a given song is streamed. To motivate our construction, we now plot sample paths of both the daily streams and chart positions of 12 different songs in Figure~\ref{fig:sample_paths}.  In a way that is similar to the plot of all top-ranked songs in Figure~\ref{fig:top_songs}, we can notice that two of the most striking features of these sample paths are jumps upward and decay downward. While some songs also have periods of relatively gradual growth, the increase of daily stream count appears to happen much more rapidly than the decrease does.  When a song does lose listens from day to day, it appears to happen slowly. There do not appear to be downward jumps like the surges in new listens that can be seen in some form in each of the sample paths. We will thus take upward jumps and downward decay as the salient features we want to capture in a model of a song's streams.

\subsection{Defining the Point Process Model}

Let us now define the model. For a given track, we let $\{N(t) \mid t \geq 0\}$ be the point process for the number of times the song has been played up to time $t \geq 0$. Without loss of generality we will suppose that the song debuts at time 0 and that there were no previous streams, i.e. $N(0) = 0$. We will furthermore suppose that $N(t)$ is a non-stationary Poisson process with rate $\lambda(t)$ defined
\begin{align}
\lambda(t) = \lambda + \beta_0 \theta_0 e^{-\beta_0 t} + \beta_1 \theta_1 e^{-\beta_1 (t-a)} \mathbf{1}_{t\geq a}
,
\label{lambdaDef}
\end{align}
where $\lambda \geq 0$, $\theta_0 > 0$, $\theta_1 > 0$, $\beta_0 > 0$, $\beta_1 > 0$, and $a \geq 0$. For an introductory overview of probability models of this type, see e.g. Section 5.4 of \citet{ross2014introduction}. This non-stationary arrival rate is precisely how we replicate the jump-and-decay structure we see, which can be seen by inspecting the parameters. First, $\lambda$ can be thought of as a baseline streaming rate, meaning the rate at which the song is played long into its life time once its hype has essentially ended. Then, because $\lambda(0) = \lambda + \beta \theta_0$, $\beta_0 \theta_0$ is the jump in streaming rate that a song gets from the initial excitement at its debut. As time progresses this jump size decays exponentially at rate $\beta_0$, which mimics the gradual decrease seen in Figure~\ref{fig:sample_paths}. Then, $\beta_1 \theta_1$ captures the size of the jump in streams that are brought to the song due to some later, external event that occurs at time $a$, and this affect then decays exponentially at rate $\beta_1$. In Subsection~\ref{subsecNumberOnes} we saw that these subsequent jumps could be caused by releases of the song's music video or of the album containing the single, but these need not be the only reasons a song's stream count could jump. In fact, the song ``Bad and Boujee'', shown in the second panel of Figure~\ref{fig:sample_paths} is an excellent example of this, as it is known to have jump up by 243\% of its previous daily stream totals following mention of the song by Donald Glover in his January 8, 2017 Golden Globes acceptance speech, per \citet{weiner2017donald}. This affect was then further amplified when \textit{Culture}, the album containing ``Bad and Boujee'', was released on January 27, 2017.

We note that this model could be easily generalized to feature multiple subsequent jumps by adding additional terms to the definition in Equation~\ref{lambdaDef}. Furthermore, let us note this is an intentionally simple model. While pure jumps and exponential decay may apply to many scenarios, sample paths with gradual growth or non-exponential decay are not represented by this model. Additionally, other observed dynamics such as the weekly periodicity are not addressed. While properly modeling these features presents an intriguing direction of future research, in this work we will adhere to the simple model and use both it and its fit as tools to further analyze this data. As we will see in Section~\ref{secCluster}, by also including the fit of the model in our analysis, we are able to implicitly address dynamics seen in the data that are not included in the model. We can also note that the model we have described is a continuous time stochastic process, yet this data set is recorded over discrete time intervals. This is not an issue; we can easily align the model and data as follows. Let $\Delta$ be the length of a day as measured in the time units upon which $N(t)$ is defined. Then, we define $N_i$ as the number of streams on the $i^\text{th}$ day, i.e. $N_i = N(\Delta i) - N(\Delta(i-1))$ for $i \in \mathbb{Z}^+$. By consequence, $N_i \sim \mathrm{Pois}(\int_{\Delta (i-1)}^{\Delta i} \lambda(t) \mathrm{d}t)$. Using this definition, we now the resulting maximum likelihood estimation procedure in Subsection~\ref{subsecFit}.

\subsection{Fitting the Model to Data}\label{subsecFit}

To fit the data, let us now describe the maximum likelihood estimate procedure for this stochastic process model. For a given song, we let $n_i$ for $i \in \{1, \dots, T\}$ be the number of streams received on the song's $i^\text{th}$ day, where $T \in \mathbb{Z}^+$ is the total number of days observed.  We will assume that the time of the subsequent jump $a$ is known and that it is a multiple of $\Delta$. Then, the parameters we want to estimate are the decay rates $\beta_0$ and $\beta_1$, the jump size coefficients $\theta_0$ and $\theta_1$, and the baseline stream rate $\lambda$. For simplicity, we will denote this full parameter set as $\boldsymbol{\Pi}$. Then, if these parameters were known, we could express the probability of observing the song's full streaming data sequence as
$$
\PP{\bigcup_{i=1}^T \{N_i = n_i\} \,\Bigg|\, \boldsymbol{\Pi}}
=
\prod_{i=1}^T \PP{N_i = n_i \mid \boldsymbol{\Pi}}
=
\frac{1}{n_i !}
\left( \int_{\Delta(i-1)}^{\Delta i} \lambda(t) \mathrm{d}t\right)^{n_i}
e^{-\int_{\Delta(i-1)}^{\Delta i} \lambda(t) \mathrm{d}t}
,
$$
where we have used the independent increments property of Poisson processes to separate the probability of the full sequence into the product of individual observation probabilities. We can likewise express the likelihood of the parameters $L(\boldsymbol{\Pi}) = L(\beta_0, \beta_1,\theta_0,\theta_1, \lambda)$ as a product of individual likelihoods, i.e. $L(\boldsymbol{\Pi}) = \prod_{i=1}^T L_i(\boldsymbol{\Pi})$ where $L_i(\boldsymbol{\Pi}) = L_i(\beta_0, \beta_1,\theta_0,\theta_1, \lambda) = \PP{N_i = n_i \mid \boldsymbol{\Pi}}$. Recalling the definition of $\lambda(t)$ from Equation~\ref{lambdaDef}, we can see that
$$
 \int_{\Delta(i-1)}^{\Delta i} \lambda(t) \mathrm{d}t
 =
 \lambda \Delta
 +
 \theta_0 \left( e^{-\beta_0 \Delta (i - 1)} - e^{-\beta_0 \Delta i} \right)
 +
 \theta_1 \left( e^{-\beta_1 (\Delta (i - 1) - a)^+} - e^{-\beta_1 (\Delta i - a)^+} \right)
 ,
$$
where $(x)^+  = \max(x, 0)$ is the positive part of $x$. Following standard techniques we will leverage the fact that the natural logarithm is a non-decreasing function and maximize the log-likelihood instead of the likelihood. For $\mathcal{L}(\boldsymbol{\Pi}) = \log\left(L(\boldsymbol{\Pi})\right)$,
% and $\mathcal{L}_i(\boldsymbol{\Pi}) = \log\left(L_i(\boldsymbol{\Pi})\right)$
 we have
\begin{align*}
\mathcal{L}(\boldsymbol{\Pi})
&=
%\sum_{i=1}^T \mathcal{L}_i(\boldsymbol{\Pi})
%=
\sum_{i=1}^T
\Bigg(
n_i \log\left(  \lambda \Delta
 +
 \theta_0 \left( e^{-\beta_0 \Delta (i - 1)} - e^{-\beta_0 \Delta i} \right)
 +
 \theta_1 \left( e^{-\beta_1 (\Delta (i - 1) - a)^+} - e^{-\beta_1 (\Delta i - a)^+} \right)\right)
 \\
 &
 \quad
-
\log \left(n_i !\right)
-
  \lambda \Delta
 -
 \theta_0 \left( e^{-\beta_0 \Delta (i - 1)} - e^{-\beta_0 \Delta i} \right)
 -
 \theta_1 \left( e^{-\beta_1 (\Delta (i - 1) - a)^+} - e^{-\beta_1 (\Delta i - a)^+} \right)
 \Bigg)
 %%%%%%%%%%%%%%%%%%%%%%%%%%%%%%%%%
 \\
 &=
 \sum_{i=1}^T
n_i \log\left(  \lambda \Delta
 +
 \theta_0 \left( e^{-\beta_0 \Delta (i - 1)} - e^{-\beta_0 \Delta i} \right)
 +
 \theta_1 \left( e^{-\beta_1 (\Delta (i - 1) - a)^+} - e^{-\beta_1 (\Delta i - a)^+} \right)\right)
 \\
 &
 \quad
-
\sum_{i=1}^T \log \left(n_i !\right)
-
  \lambda  \Delta T
 -
 \theta_0 \left( 1 - e^{-\beta_0 \Delta T} \right)
 -
 \theta_1 \left( 1 - e^{-\beta_1 (\Delta T - a)} \right)
 .
\end{align*}
%\begin{flalign}
%    &E0_i = e^{-\beta_0\Delta i}(1-e^{-\beta_0\Delta})\\
%    &E1_i = e^{-\beta_1\Delta i}(1-e^{-\beta_1\Delta})\\
%    &B0_i = \theta_0 \\
%    &B1_i = \theta_1 e^{\beta_1 a} \one_{i\geq \frac{a}{\Delta}}\\
%    &D_i  = \int_{\Delta i}^{\Delta (i+1)}\lambda(t)dt = \lambda\Delta + B_0*E_0 +  B_1*E_1\\
%    &L_i(\beta_0, \beta_1,\theta_0,\theta_1, \lambda, a) = P(N_i = n_i) = \frac{1}{n_i!}e^{-\int_{\Delta i}^{\Delta (i+1)}\lambda(t)dt}(\int_{\Delta i}^{\Delta (i+1)}\lambda(t)dt)^{n_i}\\
%    & => \frac{1}{n_i!}e^{-D_i}(D_i)^{n_i}
%\end{flalign}
Then, the partial derivatives of the log-likelihood with respect to each of the parameters are given by
\begin{align*}
\frac{\partial \mathcal{L}}{\partial \beta_0} &=
\sum_{i=1}^T \frac{n_i \theta_0}{D_i} \left( \Delta i e^{-\beta_0 \Delta i}  -   \Delta (i - 1) e^{-\beta_0 \Delta (i - 1)}  \right)
 -
 \theta_0 \Delta T e^{-\beta_0 \Delta T}
,\\
\frac{\partial \mathcal{L}}{\partial \beta_1} &=
\sum_{i=1}^T \frac{n_i \theta_1}{D_i} \left( (\Delta i - a)^+ e^{-\beta_1 (\Delta i - a)^+}  -  (\Delta (i - 1) - a)^+ e^{-\beta_1 (\Delta (i - 1) - a)^+} \right)
 -
 \theta_1 (\Delta T - a)e^{-\beta_1 (\Delta T - a)}
,\\
\frac{\partial \mathcal{L}}{\partial \theta_0} &=
\sum_{i=1}^T \frac{n_i}{D_i} \left( e^{-\beta_0 \Delta (i - 1)} - e^{-\beta_0 \Delta i} \right)
 -
 \left( 1 - e^{-\beta_0 \Delta T} \right)
,\\
\frac{\partial \mathcal{L}}{\partial \theta_1} &=
\sum_{i=1}^T \frac{n_i}{D_i} \left( e^{-\beta_1 (\Delta (i - 1) - a)^+} - e^{-\beta_1 (\Delta i - a)^+} \right)
 -
 \left( 1 - e^{-\beta_1 (\Delta T - a)} \right)
,\\
\frac{\partial \mathcal{L}}{\partial \lambda} &=
\sum_{i=1}^T \frac{n_i \Delta}{D_i} - \Delta T
 ,
\end{align*}
where $D_i = \lambda \Delta
 +
 \theta_0 ( e^{-\beta_0 \Delta (i - 1)} - e^{-\beta_0 \Delta i} )
 +
 \theta_1 ( e^{-\beta_1 (\Delta (i - 1) - a)^+} - e^{-\beta_1 (\Delta i - a)^+} )$. Likewise, the second derivatives are given by
\begin{align*}
\frac{\partial^2 \mathcal{L}}{\partial \beta_0^2} &=
 \theta_0 \Delta^2 T^2 e^{-\beta_0 \Delta T}
-
\sum_{i=1}^T \frac{n_i \theta_0}{D_i} \left( \Delta^2 i^2 e^{-\beta_0 \Delta i}  -   \Delta^2 (i - 1)^2 e^{-\beta_0 \Delta (i - 1)}  \right)
\\
&\quad
-
\sum_{i=1}^T \frac{n_i \theta_0^2}{D_i^2}
\left( \Delta i e^{-\beta_0 \Delta i}  -   \Delta (i - 1) e^{-\beta_0 \Delta (i - 1)}  \right)^2
,\\
\frac{\partial \mathcal{L}}{\partial \beta_1} &=
 \theta_1 (\Delta T - a)^2 e^{-\beta_1 (\Delta T - a)}
 -
\sum_{i=1}^T \frac{n_i \theta_1}{D_i} \left( \left((\Delta i - a)^+\right)^2 e^{-\beta_1 (\Delta i - a)^+}  -  \left((\Delta (i - 1) - a)^+\right)^2 e^{-\beta_1 (\Delta (i - 1) - a)^+} \right)
\\
&\quad
-
\sum_{i=1}^T \frac{n_i \theta_1^2}{D_i^2} \left( (\Delta i - a)^+ e^{-\beta_1 (\Delta i - a)^+}  -  (\Delta (i - 1) - a)^+ e^{-\beta_1 (\Delta (i - 1) - a)^+} \right)^2
,\\
\frac{\partial^2 \mathcal{L}}{\partial \theta_0^2} &=
-\sum_{i=1}^T \frac{n_i}{D_i^2} \left( e^{-\beta_0 \Delta (i - 1)} - e^{-\beta_0 \Delta i} \right)^2
,\\
\frac{\partial^2 \mathcal{L}}{\partial \theta_1^2} &=
-\sum_{i=1}^T \frac{n_i}{D_i^2} \left( e^{-\beta_1 (\Delta (i - 1) - a)^+} - e^{-\beta_1 (\Delta i - a)^+} \right)^2
,\\
\frac{\partial^2 \mathcal{L}}{\partial \lambda^2} &=
-\sum_{i=1}^T \frac{n_i \Delta^2}{D_i^2}
 ,
\end{align*}
and it can be observed that each of these is negative. Thus, deriving the maximum likelihood estimate is a convex maximization problem and can be solved using the convex optimization software of one's choosing, such as \citet{grant2008cvx}.

We can also note that in the special sub-case when the song's streaming decay rates are the only quantities to be estimated and are furthermore assumed to be equal, i.e. $\beta \equiv \beta_0 = \beta_1$, then one can leverage linear regression to produce a simple estimation technique. By taking the logarithm of the daily streaming data, the resulting linear trend is then equal to the true decay rate according to the non-stationary Poisson process model. Hence by the Gauss-Markov theorem, the slope of the resulting trend line is the best linear unbiased estimate of the decay rate $\beta$, see e.g. Chapter 14 of \citet{barreto2005introductory}. It is worth noting that this although this special case of estimating $\beta$ only may actually be quite informative, and this decay rate is arguably the most relevant parameter in practical application.  Because this parameter represents the rate at which the song's streams decrease from day to day, $\beta$ captures an inherent notion of long-term popularity and longevity of the song, and these are significant drivers of its success in terms of both commercial earnings and cultural relevance. For these reasons, the streaming decay rate will be central to the model-based analysis we perform in Section~\ref{secCluster}, in which we find insights for the songs contained in this data set.

\section{Data Insights via the Stochastic Process Model}\label{secCluster}

In this section, we apply the point process model that we have defined in Section~\ref{secModel} to the Spotify Top 200 data set. Through fitting this model to the data, we are able to uncover relationships between different songs in the Top 200. As a way of comparing songs, our analysis focuses on the decay rate of the daily streaming count. We will use the assumption that there is a single decay rate, i.e. $\beta = \beta_0 = \beta_1$. In the course of our analysis we find that this simple case still delivers powerful insights. In Subsection~\ref{subsecFitEval} we inspect the fit of this parameter to the data and investigate the strengths and weaknesses of the model. Then, in Subsection~\ref{subsecKMeans} we use both the decay rate and its fit to group similar songs together through use of the $k$-means clustering algorithm. In reviewing these clusters we find that the algorithm identifies songs with explainable similarities, including connections that extend beyond the scope of this data set.

\subsection{Evaluating the Fit of the Model}\label{subsecFitEval}

\begin{figure}[t]
    \centering
    \includegraphics[width = 0.49\textwidth]{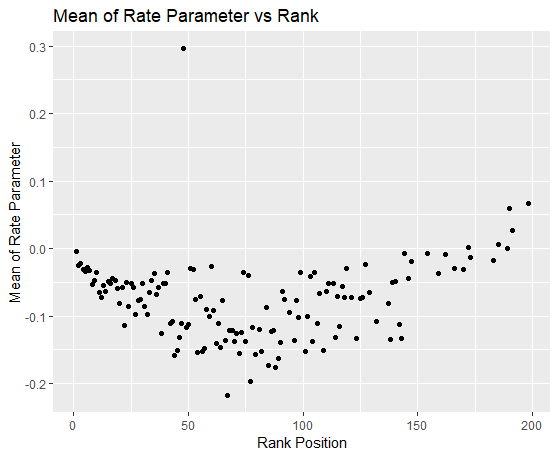}
    ~\includegraphics[width = 0.49\textwidth]{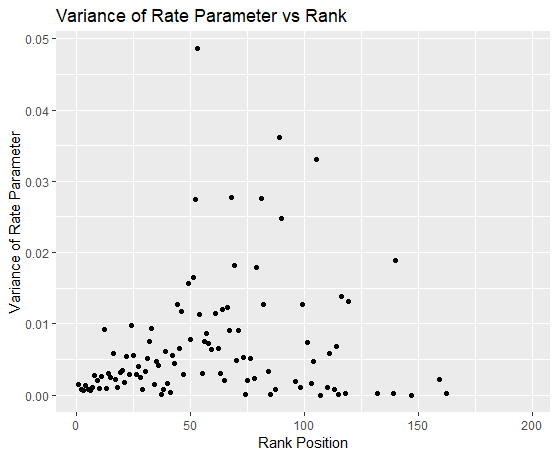}
    \caption{Comparison of each song's estimated decay rate and the highest rank it achieved.}
    \label{fig:Rate Paramter vs Rank}
    \end{figure}

    To begin our inspection of the decay rate parameter fit to each song in this data, let us first look at how it relates to the rank of the song in the Top 200. In Figure~\ref{fig:Rate Paramter vs Rank}, we have two scatter plots comparing the decay rate and the highest rank the song reached in the charts. On the left of Figure~\ref{fig:Rate Paramter vs Rank}, we take the average of the parameters of all the songs that peaked at the given rank, and on the right of Figure~\ref{fig:Rate Paramter vs Rank} we calculate the variance of the parameters across these songs. In this chart the negative value of the decay rate parameter corresponds to true negative decay, i.e. $\beta > 0$ in $e^{-\beta t}$.
    Figure \ref{fig:Rate Paramter vs Rank} shows that the middle ranks of 30 to 130 have the largest decay rates on average. The top ten ranks and ranks 150-200 have mean rate parameter close to 0. Interestingly, we also see that variance in rate parameter is greatest also for the middle ranks and lower for the top ten ranks and the bottom 50 ranks. In the case of the bottom 50 positions, we can recall that Figure~\ref{fig:staying_time_decay} showed us that these songs do not remain on the chart for very long. Thus, there could be a missing data issue here as we are not able to observe the decay of the songs streams over many days. Regardless, we are able to observe that the decay rates of middle ranks are appear both larger and less stable in comparison to the decay of the top ranked songs. This matches the intuition provided by Figure~\ref{fig:staying_time_decay}, as bigger hits are known to last longer on the charts.

    % \begin{figure}[t]
    %     \centering
    %     \includegraphics[width = 0.5\textwidth]{fit_figures/variance_rate_parameter_vs_rank.png}
    %     \caption{Variance of Rate Parameter vs Rank}
    %     \label{fig:Variance of Rate Parameter vs Rank}
    % \end{figure}

%\subsection{Rate Parameters for Top 200 Ranks}
%See Figure \ref{fig:Rate_Parameters},
%    \begin{figure}[t]
%        \centering
%        \includegraphics[width = 0.4\textwidth]{fit_figures/Dashboard_Rank_Top_10.png}
%        \includegraphics[width = 0.4\textwidth]{fit_figures/Dashboard_12-198.png}
%        \label{fig:Rate_Parameters}
%        \caption{Left: Rate Parameter for Top 10 Ranks; Right: Rate Parameter for Bottom 190 Ranks}
%    \end{figure}
    % \begin{figure}[t]
    %     \centering

    %     \label{fig:Rate_Parameter_for_Bottom_190_Ranks}
    %     \caption{}
    % \end{figure}

%\subsection{Simple Model}
%First, we try to describe the curves using a vanilla model: $y = \alpha e^{-\beta t}$. The parameter distribution of $\alpha$ and $\beta$ for some songs that have reached rank 1 in 2017 or 2018 is shown at Figure \ref{fig:simple_param_dist}.
%The negated decay rate $\beta$ is concentrated between 0.003 - 0.015. \textit{HUMBLE} is an outlier in this set of parameters.

%\begin{figure}[t]
%    \centering
%    \includegraphics[width = \textwidth]{fit_figures/params.png}
%    \caption{parameter distribution of $\alpha$ and $\beta$. All songs in the graph were at rank 1 at some point.}
%    \label{fig:simple_param_dist}
%\end{figure}

%\subsection{Fitting the Model to Data}

In addition to inspecting the fitted decay rates across the whole data set, it is important that we also investigate the fit on individual songs and see if the model matches the data.
As we will now show, this simple model works relatively well for many songs; however, some characteristics of certain songs cannot be captured by such simplicity. In particular, in some cases the decay rate of the streams is significantly slower than an exponential function. This will be especially apparent in the log-scale graph. It is worth noting that there are also songs that do not exhibit this decay slow down behavior and appear to be well represented by exponential function.

\begin{figure}[t]
    \centering
    \includegraphics[width = 0.32\textwidth]{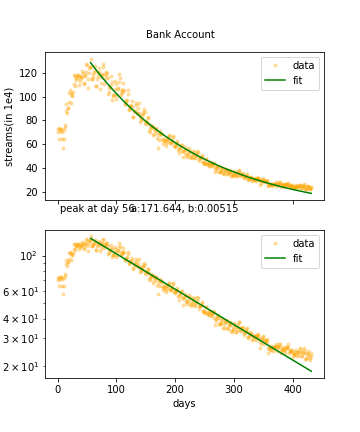}
\includegraphics[width = 0.32\textwidth]{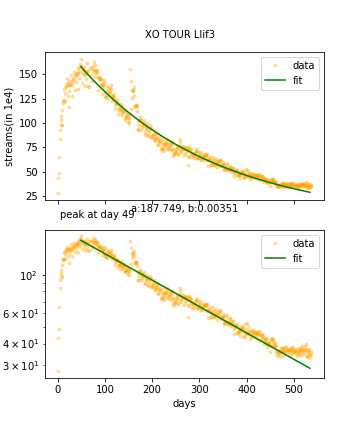}
\includegraphics[width = 0.32\textwidth]{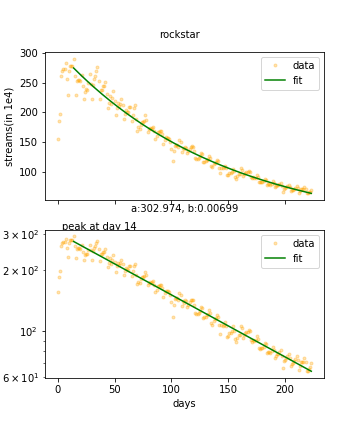}
    \caption{Comparison of data and estimated model for three songs with visually good fit.}
    \label{fig:great_fit_tracks}
\end{figure}

Let us begin our individual song model inspection by reviewing examples of the latter. In Figure \ref{fig:great_fit_tracks}, we plot the daily streaming data and the fit decay rates for three example songs, ``Bank Account'' by 21 Savage, ``XO TOUR Llif3'' by Lil Uzi Vert, and ``rockstar'' by Post Malone feat. 21 Savage. On the top row of plots we compare the data and the model on a linear-scale y-axis, and on the bottom row we plot the same curves in log-scale. In each of the figures, we plot the exponential decay from the time of the song's final peak onwards. From reviewing these, we can see that each estimated curve closely resembles the data from the Spotify Top 200.

\begin{figure}[t]
    \centering
\includegraphics[width = 0.32\textwidth]{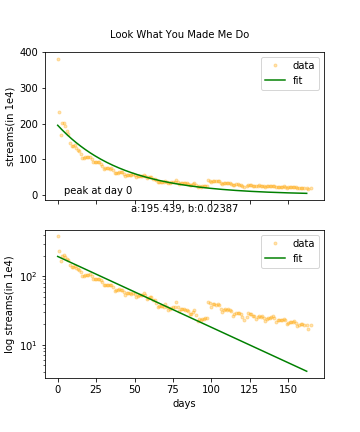}
\includegraphics[width = 0.32\textwidth]{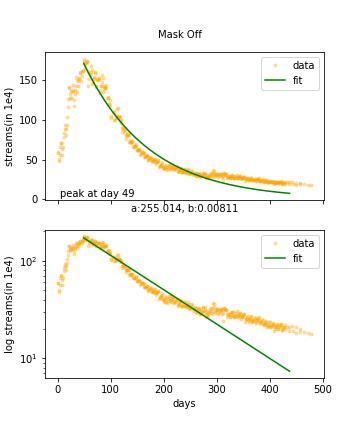}
\includegraphics[width = 0.32\textwidth]{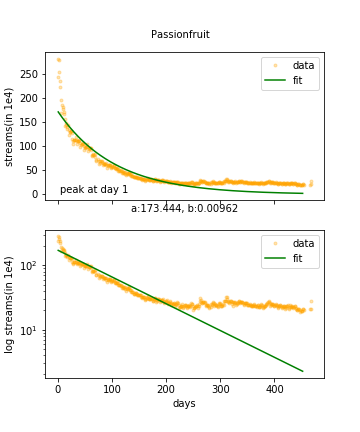}
    \caption{Comparison of data and estimated model for three songs with visually inaccurate fit.}
    \label{fig:no_jump_tracks}
\end{figure}

Now in Figure~\ref{fig:no_jump_tracks} we plot the same comparisons in both linear- and log-scale for three more songs, ``Look What You Made Me Do'' by Taylor Swift, ``Mask Off'' by Future, and ``Passionfruit'' by Drake. As is particularly noticeable in the log-scale plots, the tails of these streaming data examples are certainly heavier than the exponential curves shown. This shows one potential pitfall of the simplicity of this model and of only considering the decay rate of the song within it. For example for ``Mask Off,'' it seems as if this data may simply be better fit by a function with a tail heavier than $e^{-\beta t}$. In the case of ``Passionfruit,'' it appears that the streams may be leveling off to an underlying baseline streaming rate, which can be modeled as the constant $\lambda$ in Equation~\ref{lambdaDef}. This raises the concern that it could be possible that the better fits in Figure~\ref{fig:great_fit_tracks} could simply be a result of having observed too few data points. However, both ``XO TOUR Llif3'' and ``Bank Account'' have been observed as having stable decay rates for more than 400 days, before which each of the examples in Figure~\ref{fig:no_jump_tracks}.

While seeing these less accurate estimates is certainly a prompt for the future development of more sophisticated stochastic process models of streaming data, this need not be a barrier to finding deeper insights from the simple exponential models. In fact, in Subsection~\ref{subsecKMeans} we find that we actually use inaccuracy as a tool. By including a measure of the model fit with the decay rate for use of a clustering algorithm, we are able to recognize interesting similarities between songs.

\subsection{Finding Similar Songs through $k$-Means Clustering Analysis}\label{subsecKMeans}

As our final analysis in this paper, we now apply the well-known $k$-means clustering algorithm to the Spotify Top 200 data set. We base this clustering approach on the point process model defined in Section~\ref{secModel}. Specifically, the two quantities we use are the estimated decay rates and the accuracies.  We will use the simple estimation approach based on obtain the best linear unbiased estimate via linear regression, as described at the end of Subsection~\ref{subsecFit}, and thus we will take the resulting $R^2$ value as the measurement of the fit. Moreover, we estimate these exponential decay rates through the full data and disregard and jumps beyond time 0. Because knowing that an exponential curve doesn't nicely fit the data still tells us something about the data, we actually use the potential over-simplicity of this exponential model as a feature. In this way, this clustering procedure still yields insightful results despite the simplicity of the estimation procedure.

\begin{figure}[t]
\centering
\includegraphics[width = \textwidth]{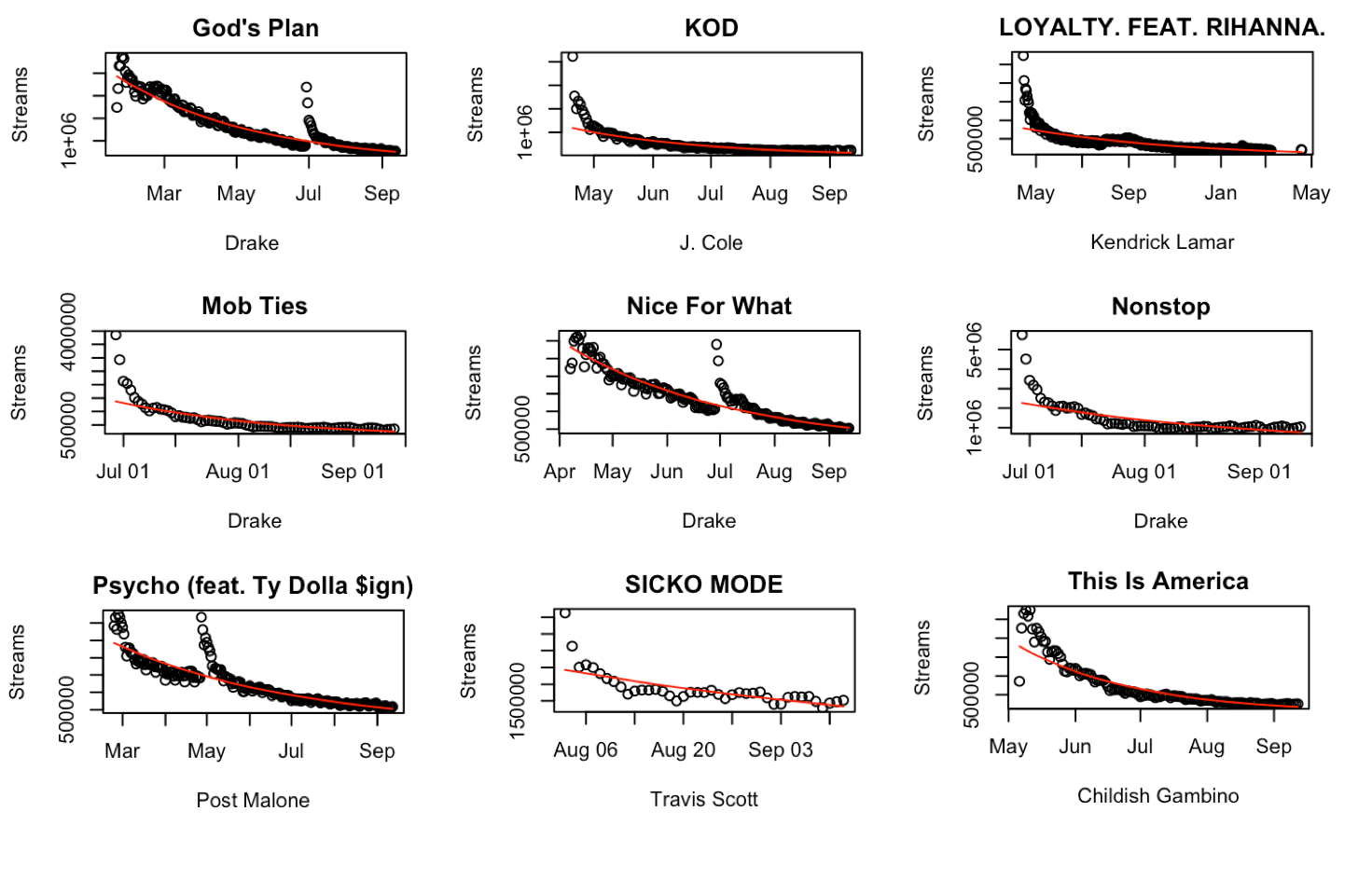}
\caption{A cluster containing major hit songs, including 5 Billboard \#1 hits.}\label{fig:cluster1}
\end{figure}

\begin{figure}[t]
\centering
\includegraphics[width = \textwidth]{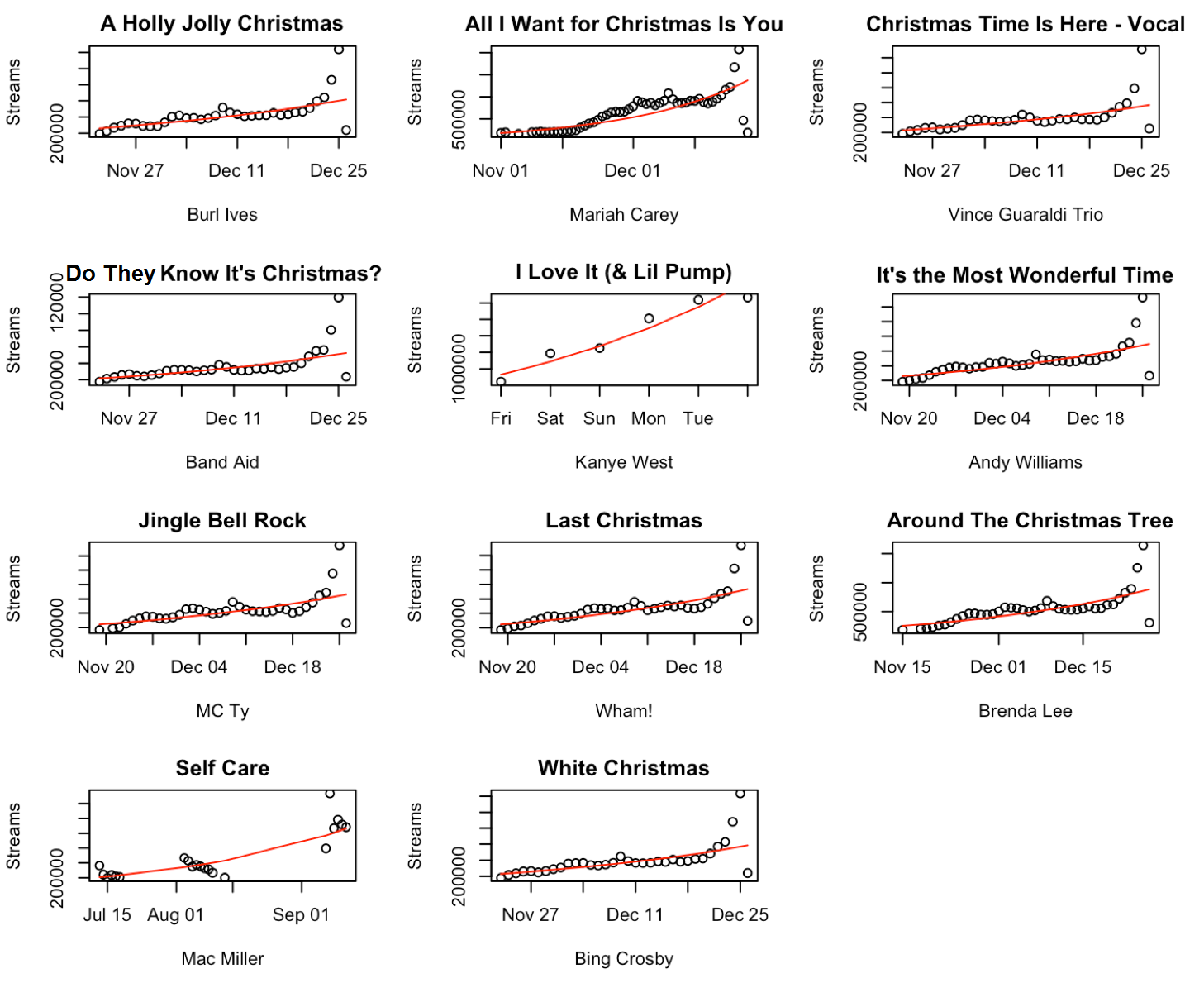}
\caption{A cluster containing primarily popular Christmas songs, as well as 2 notable outliers.}\label{fig:cluster2}
\end{figure}

\begin{figure}[t]
\centering
\includegraphics[width = \textwidth]{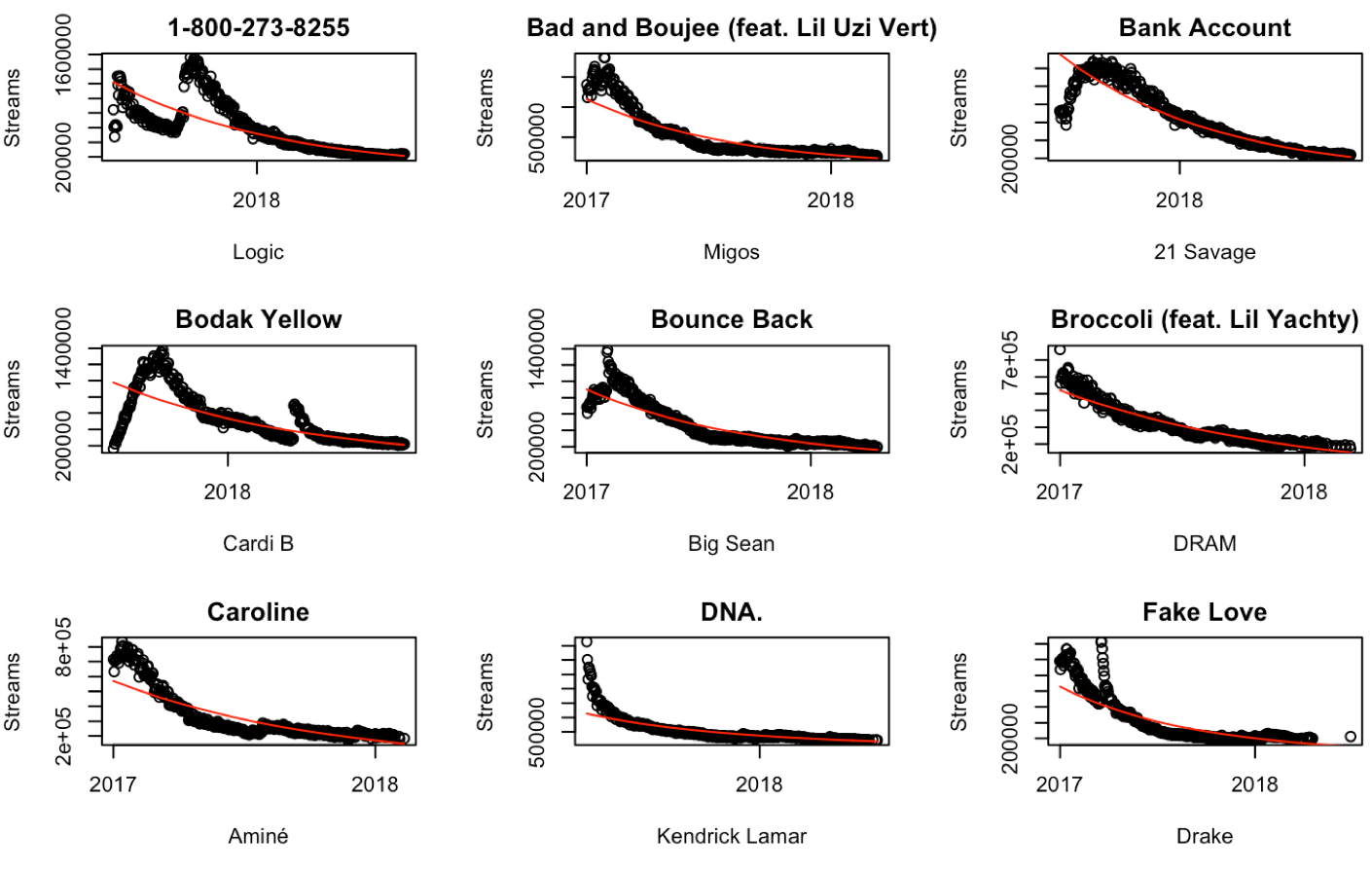}
\caption{A cluster largely comprised of songs with steady and gradual initial growth.}\label{fig:cluster3}
\end{figure}

It is also worth noting that the practical relevance of similarity among the songs' exponential decay rates can be motivated from the context of Spotify's streaming service. We believe that the similarity of streaming behaviors of two songs could be from similarity of audiences, i.e. the same types of songs are listened to by the same groups of people. Because these fans may exhibit the same listening habits across multiple songs, the recorded daily stream data may follow similar patterns. As we will see in three example clusters produced by this algorithm shown in Figures~\ref{fig:cluster1},~\ref{fig:cluster2}, and~\ref{fig:cluster3}, there are observable commonalities between these songs that transcend the calculated $\beta$ and $R^2$ values used to group them.

In our first example cluster, shown in Figure~\ref{fig:cluster1}, we find nine songs that include five of the twelve 2018 Billboard number one singles - ``God's Plan,'' ``Nice for What,'' ``This is America,'' ``Psycho,'' and ``Sicko Mode.'' Fascinatingly, ``Sicko Mode'' wouldn't top the charts until three months after this data set ended, a powerful connection identified by this clustering algorithm and this model. It is also worth recalling that the Billboard Hot 100 rankings draw on much more than just the Spotify streaming data, including radio play, physical sales, and digital sales. The remaining four songs feature two of the most popular eventual singles from Drake's \textit{Scorpion}, ``Mob Ties'' and ``Nonstop,'' a song from J. Cole that broke the Spotify opening day U.S. streams record, ``KOD,'' and the Kendrick Lamar and Rihanna hit, ``LOYALTY.'' Thus, in this cluster the algorithm has recognized songs of extreme popularity simply from the estimates of their streaming decay rates and the overall empirical fit of those estimates, but not the actual daily streaming counts themselves.

\begin{figure}[H]
\centering
\includegraphics[width = \textwidth]{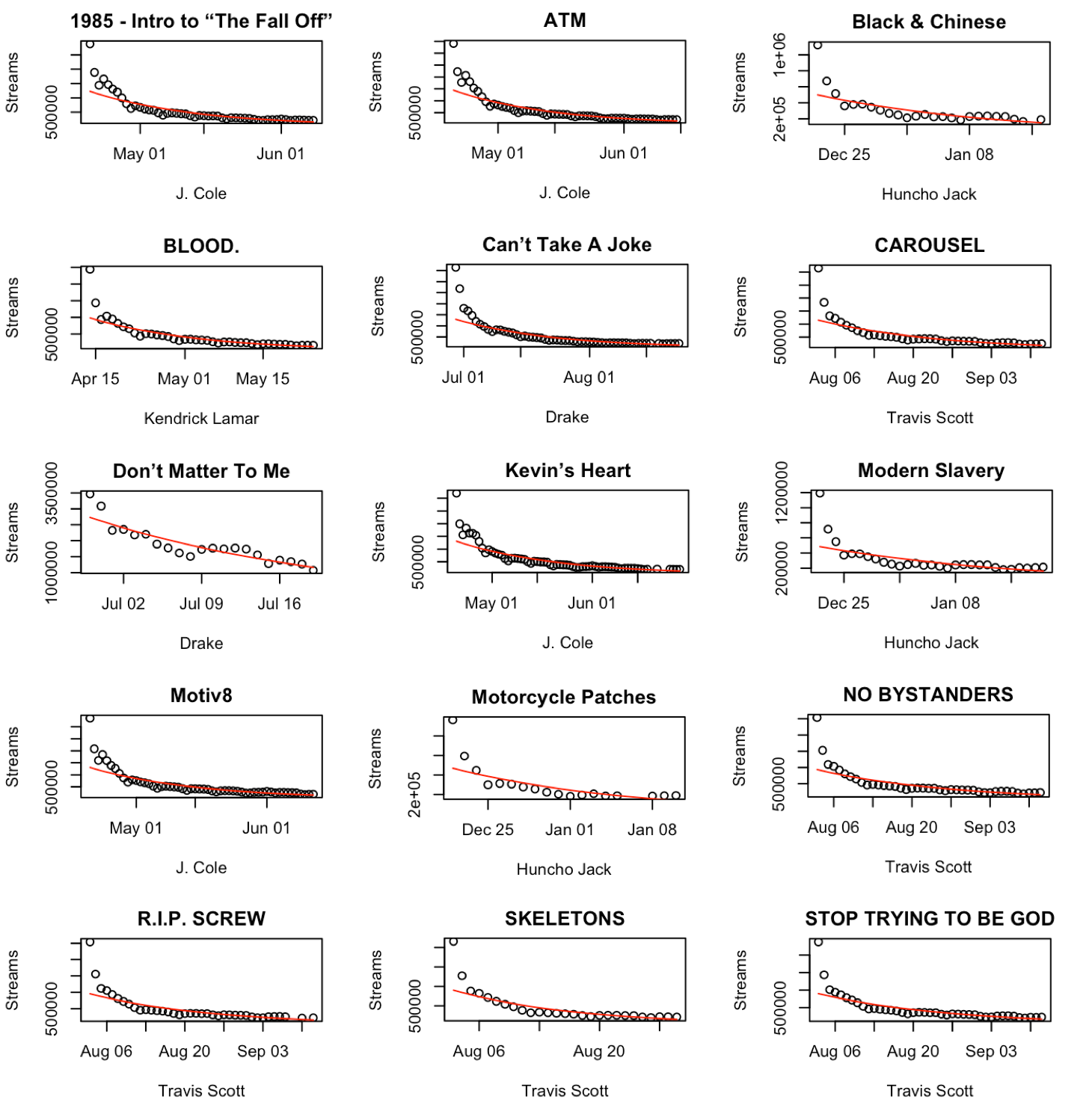}
\caption{A cluster containing major hit songs, including 5 Billboard \#1 hits.}\label{fig:cluster4}
\end{figure}

By comparison to the collection of mega-hits in Figure~\ref{fig:cluster1}, Figure~\ref{fig:cluster4} contains a more modest collection of songs.  This cluster contains fifteen songs from five recent smash hit hip-hop albums that weren't quite the records' top hits but were still generally popular. In this way, this group can be thought of as the legacy cluster, as these songs are likely listened to because of the prestige of the albums containing them and the larger singles neighboring them. To demonstrate this point, five of these songs are from Travis Scott's Billboard topping \textit{ASTROWORLD}, but that album's number one single ``Sicko Mode'' isn't on this list; it is instead with the hits in Figure~\ref{fig:cluster1}. Since subsets of these songs were likely listened to in unison, Figure~\ref{fig:cluster4} motivates the idea of the similar decay rates capturing similar groups of listeners. This idea is re-enforced by the fact that all these songs share a genre, which again suggests a potential overlap in fans.

Figure~\ref{fig:cluster2} contains what could be affectionately deemed the Christmas music cluster. Perhaps most notable among these is Mariah Carey's classic ``All I Want for Christmas Is You,'' which is actually also a Spotify Top 200 number one hit and can be seen as brief blip in the middle of Figure~\ref{fig:top_songs}. These holidays songs all have an interesting behavior pattern. Unlike most songs, they trend upwards as the user base feels increasingly festive, followed by an immediate crash back down on December 26  once the listeners are taking down the tree. Hence, the estimated exponential parameters for this group are actually for exponential growth rather than exponential decay. The two exceptions to the yuletide theme of this group are each notable, and both would likely experience a steady decline in the weeks following the end of this data set that would distinguish them from the holiday music. The first, ``I Love It'' by Kanye West and Lil Pump, is in the early stages of a viral hit in this date range, being a surprise, unorthodox single that became a top hit. The second, ``Self Care,'' is a sadder story, as the artist Mac Miller died from an overdose in September 2018. The activity seen in this data set was the song being revisited by fans following his passing.

As a demonstration of the interpretative power of clustering using the measure of fit, Figure~\ref{fig:cluster3} can be seen as a collection of the late bloomers. This cluster captured several songs that steadily accumulated hype until they became major hits, like ``Bad and Boujee'' by Migos feat. Lil Uzi Vert, ``Bank Account'' by 21 Savage, ``Bodak Yellow'' by Cardi B, ``Bounce Back'' by Big Sean, and ``Caroline'' by Amin\'e, or experienced significant jumps long after their release, such as ``1-800-273-8255'' by Logic feat. Alessia Cara and Khalid and ``Fake Love'' by Drake. As a side note, it is worth pointing out that the  large jump visible in the sample path for ``1-800-273-8255'' came after Logic's August 27, 2017 MTV Video Music Awards show performance, motivating public performances as another type of event that can drive new listens.

%Cardi B represents three of these eighteen tracks, proving that genius is never appreciated in its time.

\section{Conclusion}

%In conclusion, we have explored Spotify's music streaming service using data analysis and a point process model.

In this work, we have analyzed the dynamics of the Spotify Top 200 rankings through data. We have performed an exploratory investigation of the ranks, durations, and behaviors of the songs contained in this list. In doing so we have seen that the average number of daily streams is power-law function of the song's rank, that the highest ranks are rarely achieved, and that the more popular songs have a greater longevity than their lower ranked counterparts. Additionally, we have seen a jump and decay structure to a song's daily stream count, in which jumps are often caused by an external event bringing the song more attention. This exploration led us to define a non-stationary point process model of a song receiving streams. This model is a time-varying Poisson process with intensity designed to mimic the jumps and decays seen in the data. After defining the model we described how to estimate its parameters using the streaming data. Once we have fit this model to data, we used it to further analyze the rankings and identify similarities between the songs seen in the Spotify Top 200. In particular, we have identified groups of related songs through a $k$-means clustering approach that relies only on the estimated parameters and measures of their empirical error. In examples of this cluster, we saw that this information was enough to not only identify songs that achieved high levels of streaming on Spotify, but also to recognize songs of similar success beyond just the Spotify platform, even when that was achieved after the time range of this data set.

As an example application resulting from this work, artists, labels, and other strategic management groups could evaluate whether songs have stream count decay rates similar to other historical hits as a way of determining which songs would be good candidates for the next single off of an album. Upon this evaluation, the song could then be further promoted via an event such as a music video release, which we have seen to be capable of creating a jump in daily streams. In addition to applications of this work, there are also many opportunities for future research, both empirical and theoretical. For example, one direction we are interested in pursuing is modeling random jump sizes in the point process and determining how to estimate them on one sample path when only a handful will be observed for each song. Another data-based direction would be to study songs not on the Top 200 but rather on the long tail, thus requiring additional data. This may be particularly relevant for local and/or independent music such as what is the subject of recent research \cite{turnbull2014using,turnbull2018local,akimchuk2019evaluating} and the associated local music recommendation service Localify\footnote{See: \texttt{https://www.localify.org/}}. As predicted by \citet{anderson2004long}, the total of these more niche offerings likely comprise a large portion of the overall streaming consumption, and are thus important for artists, fans, and platforms alike. For a final example of future research directions, we can note that this paper gives rise to a new decision problem in operations research, in which one seeks to optimally time the release of albums or music videos in support of a single's commercial success. This idea likely also extends to remixes of the single. In this context it appears that Lil Nas X may have already perfected the technique. Billie Eilish also recently used similar strategies, releasing a remix to ``Bad Guy'' with Justin Bieber that helped propel her song to the number one spot. Hence we propose studying the \textit{Old Town Road Problem}, deriving an optimal external event strategy, and comparing it with the observed techniques that led to the all-time longest streak atop the Billboard Hot 100.

%\section{Acknowledgements}
%The CORAL (Cornell Operations Research Applied Learning) Team would like to extend a great thanks to Andrew Daw and Professor Jamol Pender for guiding this exciting project.

\bibliographystyle{plainnat}
\bibliography{sigproc}

% \balancecolumns
%\appendix
%
%
%
%\section{Model Description}

% That's all folks!

\end{document}